\documentclass{custom}

\usepackage{times}
\usepackage{graphicx} % when using Latex and dvips
\begin{document}

\title{Large zenith angle observations with the high-resolution GRANITE III camera}
\author[1]{D. Petry$^1$ and the VERITAS Collaboration$^2$}
\affil[1]{$^1$Dept. of Physics and Astronomy, Iowa State University, Ames, IA 50010, USA}
\affil[2]{$^2$Whipple Observatory, Amado, AZ, 85645, USA}

\correspondence{petry@iastate.edu}

\firstpage{1}
\pubyear{2001}

% \titleheight{11cm} % uncomment and adjust in case your title block
                     % does not fit into the default and minimum 7.5 cm

\maketitle

\begin{abstract}
The GRANITE III camera of the Whipple Cheren\-kov Telescope at the Fred
Lawrence Whipple Observatory on Mount Hopkins, Arizona (2300 m a.s.l.)
has the highest angular resolution of all cameras used on this
telescope so far. The central region of the camera
has  379 pixels with an individual angular diameter of 0.12$^\circ$. 
This makes the instrument
especially suitable for observations of gamma-induced air-showers
at large zenith angles since the increase in average distance 
to the shower maximum leads to smaller shower images in the
focal plane of the telescope. We examine the performance of
the telescope for observations  of 
gamma-induced  air-showers at zenith angles up to 63$^\circ$
based on observations of Mkn 421 and using
Monte Carlo Simulations. An improvement to the standard
data analysis is suggested. 
\end{abstract}

% Reminder: \citet{mc} gives XY et al. (19xx) 
%           \citep{mc} gives (XY et al., 19xx)

\vspace{-0.1cm}

\section{Introduction}

\vspace{-0.1cm}
By large-zenith-angle observations we mean targeting an astronomical
source  with a ground-based telescope when it is more than 45$^\circ$ 
away from the zenith. 
Many authors have stressed the potential of using 
Imaging Atmospheric Cheren\-kov Telescopes (CTs) for
observations of sources of high-energy
$\gamma$-rays at large  zenith angles 
(see e.g. \citet{konopelko99a} and references therein).
Several successful observations have already been made
(e.g. \citet{tanimori}, \citet{hegra}, \citet{chadwick}, 
\citet{krennrich99}, \citet{mohanty}).

Since the effective depth of the atmosphere increases
to a good approximation as $1/\cos(\vartheta)$  (where $\vartheta$
denotes the zenith angle), the distance
of the telescope  to the shower maximum increases 
proportionally\footnote{Note that this approximation begins to break down
at $\vartheta = 70^\circ$ due to the curvature of the earth.}. 
The Cheren\-kov light produced in the shower
illuminates a larger area on the ground which leads
to an increase in the effective collection area
of the CT. This increase should very roughly be 
as large as $1/\cos^2(\vartheta)$ but is in reality even larger
due to changes in the shape of the lateral distribution of the
Cherenkov photons density. The expected drawbacks of large-$\vartheta$
observations
are an increased energy threshold and possibly worse background
suppression since  the
average Cheren\-kov photon density caused by the shower
at the telescope decreases with increasing $\vartheta$ 
roughly as $\cos^2(\vartheta)$. Furthermore, the angular
extension of the shower image decreases. This means
that with a given CT photomultiplier camera, the image
is fainter and contained in fewer pixels. There is less
information for distinguishing $\gamma$-induced showers
from the hadronic background showers.

It is a priori not obvious by how much the gain in collection
area will compensate for the decreasing image quality.
A camera with  higher angular resolution should suffer less
from the effects of shrinking images. It is therefore especially 
interesting to investigate the performance of the new GRANITE III
camera on the Whipple telescope (Mt. Hopkins, Arizona) which has 
an angular resolution of 0.12$^\circ$ in its central region, the 
best resolution ever available on this CT. This camera is described
in Finley et al. (these proceedings).

In this article we derive the performance of GRANITE III
as a function of the zenith angle $\vartheta$ and  investigate
possible improvements to the present standard data analysis
which is optimized for small $\vartheta$. We do this by
examining the performance of the standard analysis at
different zenith angles using simulated $\gamma$-showers
and real telescope data from off-source observations.
The improvements are then verified using observations of
Mkn 421. It is especially helpful
that Mkn 421 was in a high state during the observations.

\vspace{-0.1cm}

\section{Monte Carlo Data}

\vspace{-0.1cm}
The Monte Carlo (MC) data used in this study was specially produced
using the simulation code described by \citet{mc}. The agreement
of this MC data with the data obtained from the Whipple telescope
in the 2000/2001 observing season was thoroughly tested. Special
measurements of the telescopes point spread function, photomultiplier
efficiency, light-cone collection efficiency and the
photoelectron-to-digital-count conversion  were made and the results 
inclu\-ded in the simulation (Krennrich et al. 2001). Realistic noise was
added to the pixel values in order to simulate the night sky
background light.

Since all shower image parameters are expected to change linearly
with the average distance to the shower maximum which is in turn expected
to scale with $1/\cos(\vartheta)$, the MC data was produced at
equidistant values of $1/\cos(\vartheta)$. Table \ref{tab-mc}
gives an overview of the production parameters. Only gamma primaries
were produced. For studies of hadronic showers, off-source observations
with the Whipple telescope were used.

\begin{table}
\caption{\label{tab-mc} Production parameters of the Monte Carlo data
for $\gamma$-induced showers used in this study. $E$ denotes the primary
energy, $r$ the impact parameter.}
\begin{tabular}{l|cccccc}
$\vartheta$ ($^\circ$)  & 20 & 40 & 48 & 54 & 58 & 62 \\ 
$1/\cos(\vartheta)$ & 1.064 & 1.3 & 1.5 & 1.7 & 1.9 & 2.1 \\
shower number & $8\cdot10^5$ & $2\cdot10^5$ & $10^5$ & $10^5$ & $10^5$ & $10^5$ \\
spectral index & 2.5 & 2.5 & 2.5 & 2.5 & 2.5 & 2.5 \\
min. $E$ (GeV)  & 50 & 100 & 180 & 200 & 200 & 200 \\
max. $E$ (TeV)  & 100 & 100  & 100  & 100   & 100  & 100 \\
max. $r$ (m) & 310 & 400 & 550 & 650 & 760 & 870 \\
\end{tabular}
\end{table}                          

\vspace{-0.1cm}

\section{Data from the observation of Mkn 421 and off-source positions}

\vspace{-0.1cm}
The data was selected on the basis of the observer logs using the
VERITAS online log sheet database. Only prime-weather data from 2000 or 2001 with no
negative comments in the log sheets was accepted. A total of
7.0 h of Mkn 421 data with  $6^\circ < \vartheta < 22^\circ$ (749801 events),
 14.4 h  of Mkn 421 data with $34^\circ < \vartheta < 64^\circ$ (1158149 events)
and 14.5 h of off-source data with $6^\circ < \vartheta < 64^\circ$ (1074715 events)
were used.

\vspace{-0.1cm}

\section{Data analysis and results}
\vspace{-0.1cm}

\subsection{Telescope performance}
\vspace{-0.1cm}

\begin{figure}
\centering 
\includegraphics[width=8.0cm]{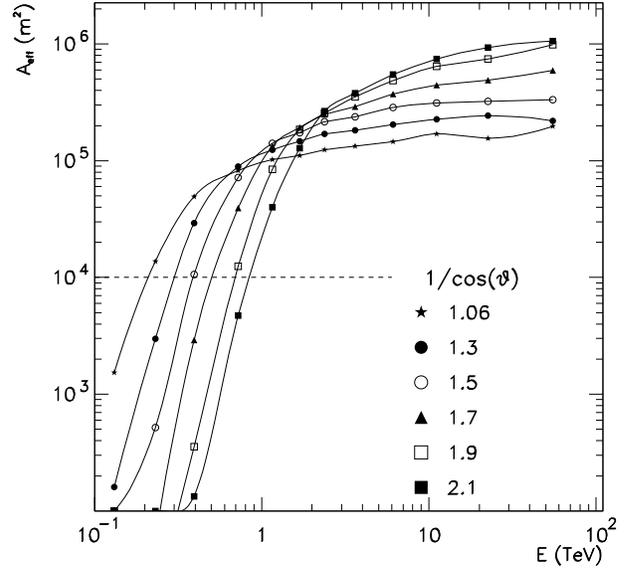}
\caption{\label{fig-efcoltrig} The effective $\gamma$-ray collection area
after trigger of the Whipple Telescope with the GRANITE III camera and the
setup of the 2000/2001 observing period. Each curve shows the collection area
of a different zenith angle ($\vartheta$) as indicated by the
symbols.}
\end{figure}

{\sloppy
The data was subjected to the standard Whipple telescope
image processing which leads to a set of 7 parameters
for each shower ({\small ALPHA, LENGTH, WIDTH,  DIST, SIZE, MAX1, MAX2}).
The cut on the signal in the two brightest pixels
of the image ({\small MAX1}$>$15 and {\small MAX2}$>$12 photoelectrons) 
essentially determines the trigger threshold for 
$\gamma$-showers in the raw data.
By applying the {\small MAX1} and {\small MAX2} cuts to the MC $\gamma$-shower dataset,
the effective collection area for primary $\gamma$-rays after trigger 
of the Whipple telescope was calculated (Fig.\,\ref{fig-efcoltrig}). 
}

\begin{figure}[h!] 
\centering
\includegraphics[width=8.0cm]{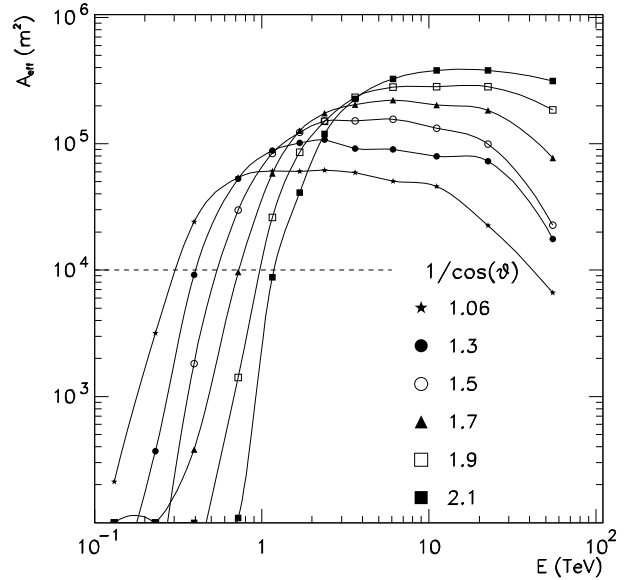}
\caption{\label{fig-efcolscuts} As Fig.\,\protect\ref{fig-efcoltrig}
but with the additional application of the Supercuts 2001
$\gamma$/hadron separation.}
\end{figure}

The suppression of the hadronic background  is
achiev\-ed by applying cuts to all image parameters mentioned above.
By optimizing on data obtained from Crab Nebula observations
at $\vartheta < 30^\circ$, a set of standard cuts was obtained
(see Finley et al., these proceedings). These so-called
Supercuts 2001  are constant cuts which are
neither functions of $\vartheta$ nor any other parameter.
By applying these cuts to the MC $\gamma$-shower dataset,
the effective collection area of the Whipple telescope
was determined (Fig.\,\ref{fig-efcolscuts}).

By convoluting the collection areas in Fig.\,\ref{fig-efcolscuts}
with a power-law spectrum with index 2.7 and determining the
energy at which the resulting rate peaks, one obtains the
``energy of peak sensitivity'' $E_p$ for each zenith angle. The values
are given in table \ref{tab-ethresh}. 

\begin{table}[hbt]
\caption{\label{tab-ethresh} Values of the energy of peak sensitivity  $E_p$
for the Whipple Telescope in the 2000/2001 observing season using Supercuts2001.}
\begin{tabular}{l|cccccc}
$1/\cos(\vartheta)$ & 1.064 & 1.3 & 1.5 & 1.7 & 1.9 & 2.1 \\
$E_p$ (GeV) & 380 & 520 & 790 & 1200 & 1500 & 2200 \\
\end{tabular}
\end{table}

\vspace{-0.1cm}

\subsection{Possible improvements of the $\gamma$/hadron separation}
\vspace{-0.1cm}

\begin{figure}[b!] 
\centering
\vspace*{-2.0mm}
\includegraphics[width=7cm]{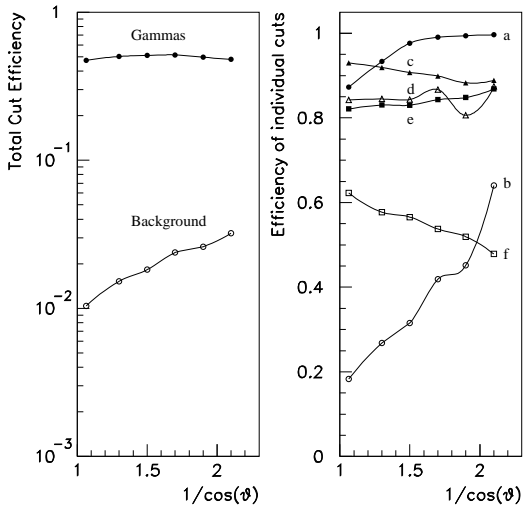}
\caption{\label{fig-effplot3} The efficiency (eff.) of the Supercuts2001
for simulated $\gamma$s (solid symbols) and real background data (open symbols).
Left: eff.  of all cuts after the MAX1 \& MAX2 cut. Right: (a, b) eff. of the upper
LENGTH and WIDTH cuts after all other cuts, (c, d) eff. of the DIST cut after all
other cuts, (e, f) eff. of the LENGTH/SIZE cut after all other cuts.}
\end{figure}

Comparing Figs. \ref{fig-efcoltrig} and \ref{fig-efcolscuts}, improvements
seem possible mainly at large energies and not so much at large zenith
angles.
However, as Fig.\,\ref{fig-effplot3} (left plot) shows, the background rejection of
the Supercuts2001 deteriorates dramatically with increasing $\vartheta$.
We examine the individual image parameters in order to identify
the reasons.

The {\small ALPHA} distribution after cuts has a shape which is
independent of $\vartheta$. The cut at 15$^\circ$ is optimal for all $\vartheta$.

The {\small DIST} cut ($0.4^\circ<${\small DIST}$<1.0^\circ$) has (after all other cuts) a
nearly $\vartheta$-independent efficiency both for $\gamma$s and background
(see Fig.\,\ref{fig-effplot3}). The peak of the {\small DIST} distribution for $\gamma$s
before cuts moves towards smaller values as $\vartheta$ increases, but for the $\gamma$s
below $0.4^\circ$ the {\small ALPHA} distribution is so broad that more than 60\% are
eliminated by the {\small ALPHA} cut (for all $\vartheta$). Also the energy resolution
for events with small {\small DIST} is poor and since the solid angle is small, they
constitute a small part of the total events. The upper {\small DIST} cut is determined
by the field of view of the camera. It is therefore not necessary to vary
the {\small DIST} cut with $\vartheta$.

\begin{figure}[t!] 
\centering
\includegraphics[width=7.5cm]{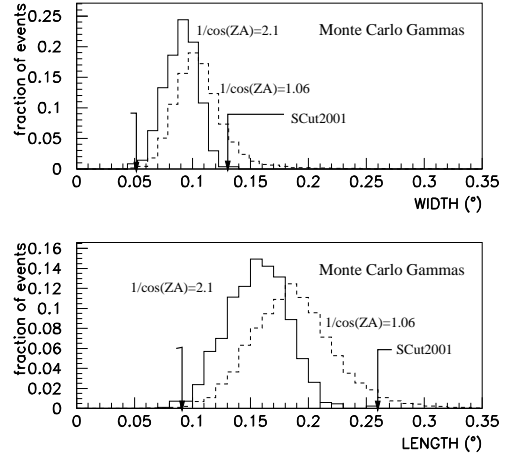}
\vspace*{-2.0mm}
\caption{\label{fig-widthlength} Comparison of the WIDTH and LENGTH
distributions (after all other cuts) 
of simulated $\gamma$ events at small and large zenith angles. See text.}
\end{figure}
\begin{figure}[h!]
\centering
\includegraphics[width=7.5cm]{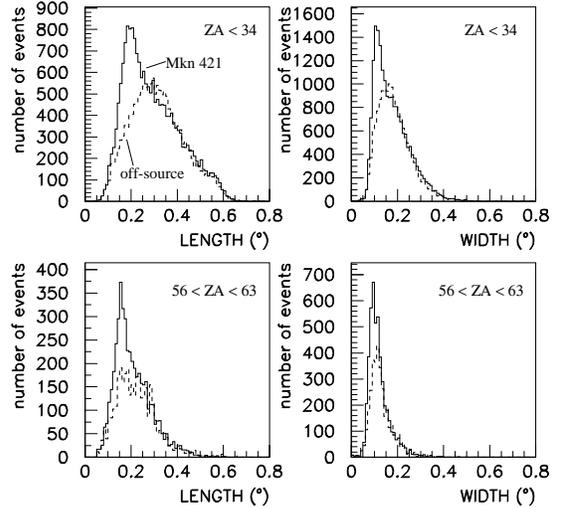}
\vspace*{-2.0mm}
\caption{\label{fig-m4widthlength} Comparison of the WIDTH and LENGTH
distributions (after all other cuts) for observations of Mkn 421 and off-source positions 
at small and large zenith angles. See text.}
\end{figure}

The lower cuts on {\small WIDTH} and {\small LENGTH}
 ({\small WIDTH}$>0.05^\circ$ and {\small LENGTH}$>0.09$)
are  dictated by the optics
of the telescope and therefore nearly independent of $\vartheta$.
This is illustrated by Figs. \ref{fig-widthlength} and \ref{fig-m4widthlength}.
While their lower end remains at the same values, the 
{\small WIDTH} and {\small LENGTH} distributions become, however, more narrow as  $\vartheta$
increases and a $\vartheta$-independent cut has too large background efficiency
(see also Fig.\,\ref{fig-effplot3}). 
The upper cuts on {\small WIDTH} and {\small LENGTH} ({\small WIDTH}$<0.13^\circ$ and
 {\small LENGTH}$<0.26$ in the
Supercuts2001) are therefore the prime candidate for improving the background rejection.

Finally, the cut on {\small LENGTH}/{\small SIZE} ( $< 1.3\times10^{-3}(^\circ$/photo\-electron))
shows  a constant $\gamma$-efficiency and a decreasing background efficiency
with increasing $\vartheta$ (see Fig.\,\ref{fig-effplot3}). Here, little improvement
seems possible. 

\begin{figure}[htb]
\centering 
\vspace{-0.2cm}
\includegraphics[width=6.4cm]{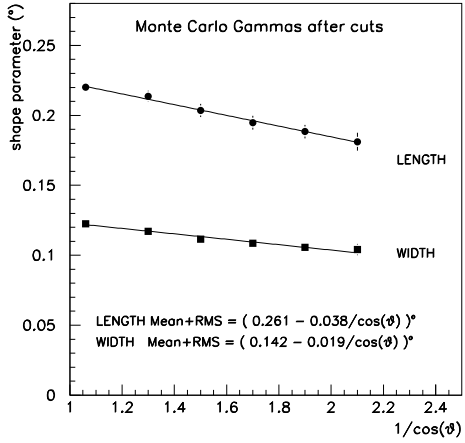}
\vspace{-0.2cm}
\caption{\label{fig-widthxvsza} The upper WIDTH and LENGTH cut values with $\gamma$-efficiency
84\% as a function of $1/\cos(\vartheta)$. The cuts are calculated
by adding one standard deviation to the mean of the distribution. See text.}
\end{figure}

In order to improve the background rejection of the upper cuts on
 {\small WIDTH} and {\small LENGTH}, we examine as a function of $\vartheta$ the cut value 
with a $\gamma$-efficiency of 84\%. Since both the {\small WIDTH} and
the {\small LENGTH} distributions for pure $\gamma$s after cuts 
are Gaussian to a very good approximation, we
determine the mean and the standard deviation of the simulated
distributions at each value of $\vartheta$ by fitting a Gaussian. 
The sum of mean and standard deviation is then plotted versus $1/\cos(\vartheta)$ 
(Fig.\,\ref{fig-widthxvsza}).

As one would expect from the geometry of the shower observation,
there is a linear dependence of the shower image size, and therefore also the
{\small WIDTH} and {\small LENGTH} values under discussion,  on $1/\cos(\vartheta)$.
The result of a linear fit to the two groups of points is given
in Fig.\,\ref{fig-widthxvsza}. In order to obtain an expression for
$\vartheta$-dependent upper cuts on {\small WIDTH} and {\small LENGTH}, we scale the two
lines such that they assume the Supercuts2001 value at the zenith angle
where the supercuts were optimized which is $\vartheta \approx 20^\circ$.
The resulting cuts are:
\begin{equation}
\label{equ-newcuts}
\begin{array}{rcl}
	\mathrm{WIDTH} & < & ( 0.15 - 0.020/\cos(\vartheta) )^\circ\\
	\mathrm{LENGTH} & < & ( 0.31 - 0.045/\cos(\vartheta) )^\circ\\
\end{array}
\end{equation}
We replace the corresponding $\vartheta$-independent cuts in the
Supercuts2001 by the cuts in equation \ref{equ-newcuts} and compare
the performance of this new set of cuts (the ``zenith-angle-dependent Supercuts'')
with that of the standard Supercuts2001.

\begin{figure}[b!]
\centering
\includegraphics[width=7cm]{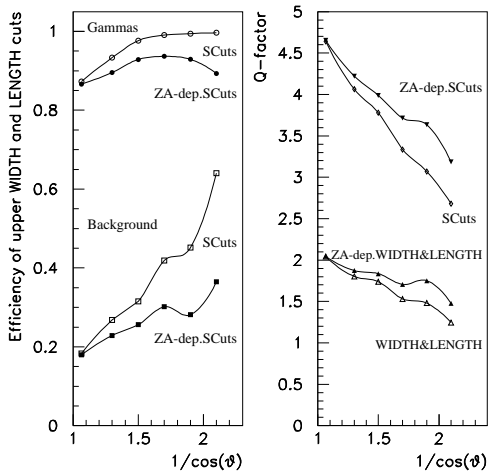}
\vspace*{-0.2cm}
\caption{\label{fig-effq}(a) (left) The efficiency of the upper cuts on
WIDTH and LENGTH after all other cuts for simulated $\gamma$s and real background data
in the case the Supercuts2001  and the zenith-angle-dependent
Supercuts derived here. (b) (right) The total quality factor of the
whole set of cuts (upper two curves) and the quality factor of only
the  upper cuts on WIDTH and LENGTH after all other cuts.}
\end{figure}
\begin{figure}[h!]
\centering
\includegraphics[width=7cm]{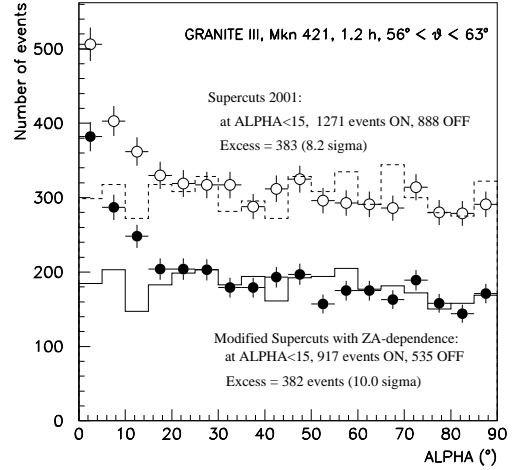}
\vspace*{-0.2cm}
\caption{\label{fig-alphademo} 
The distribution of the orientation angle ALPHA
for observations of Mkn 421 (symbols) and  off-source observations (lines)
after application of Supercuts2001 (upper curves) and  the zenith-angle-dependent
Supercuts derived here.}
\end{figure}

Fig.\,\ref{fig-effq} shows that the $\vartheta$-dependent cuts
improve the background rejection by up to a factor 2 at the largest zenith angles
while they reduce the $\gamma$-efficiency by only up to 10\%.
Furthermore, the overall quality factor ($\gamma$-efficiency divided by the square-root of
the background efficiency) drops from 4.7 to 2.7 for the  standard Supercuts2001
when going from $\vartheta = 20^\circ$ to $60^\circ$, while it drops only
to 3.2 when the $\vartheta$-dependent cuts are used. 

The improvement in sensitivity is verified using the Mkn 421 observations.
Fig.\,\ref{fig-alphademo} shows the result for the data at the largest $\vartheta$
available in this study. The expected improvement in significance from the Q-factors
in Fig.\,\ref{fig-effq} is $3.64/3.07=1.19$, the observed improvement is $10.0/8.2=1.21$ .
For intermediate zenith angles (45$^\circ$-56$^\circ$) the expected improvement is
$3.90/3.60=1.08$, in the Mkn 421 dataset we observe $15.1/14.2=1.06$. At smaller $\vartheta$
the two sets of cuts converge and no significant improvement is expected or observed.

\vspace{-0.2cm}

\section{Conclusion} 

\vspace{-0.1cm}
We have quantified the performance of the GRANITE III camera on the
Whipple Telescope and the current standard analysis at zenith angles $\vartheta < 63^\circ$ 
and identified improvements to the standard data analysis for $\vartheta > 45^\circ$. 

\vspace{-0.1cm}

\begin{acknowledgements}
The VERITAS Collaboration is supported by the U.S. Dept. of Energy,
NSF, the Smithsonian Institution, PPARC (U.K.) and
Enterprise-Ireland.

\vspace{-0.1cm}

\end{acknowledgements}

\end{document}